# Defect induced structural and thermoelectric properties of $Sb_2Te_3$ alloy

Diptasikha Das[1], K. Malik[1], A. K. Deb[2], Sandip Dhara[3], S. Bandyopadhyay[1,4] and Aritra Banerjee[1,4,a]

[1]Department of Physics, University of Calcutta, 92 A P C Road, Kolkata 700 009, India

[2]Department of Physics, Raiganj College (University College), Uttar Dinajpur 733 134, India

[3]Surface and Nanoscience Division, Indira Gandhi Centre for Atomic Research, Kalpakkam 603102, India

[4]Center for Research in Nanoscience and Nanotechnology, University of Calcutta, JD-2, Sector-III, Saltlake City, Kolkata 700 098, India

**ABSTRACT**

Structural and thermoelectric properties of metallic and semiconducting $Sb_2Te_3$ are reported. X-Ray diffraction and Raman spectroscopy studies reveal that semiconducting sample have higher defect density. Nature and origin of possible defects are highlighted. Semiconducting $Sb_2Te_3$ hosts larger numbers of defects, which act as scattering center and give rise to the increased value of resistivity, thermopower and power factor. Thermopower data indicates *p*-type nature of the synthesized samples. It is evidenced that the surface states are often mixed with the bulk state, giving rise to metallicity in $Sb_2Te_3$. Role of different scattering mechanism on the thermoelectric property of $Sb_2Te_3$ is discussed.

---

[a] Author to whom correspondence should be addressed. Electronic mail: arbphy@caluniv.ac.in

## I. INTRODUCTION

In Sb-Te system, $Sb_2Te_3$ is the most stable compound. $Sb_2Te_3$ is a well-known thermoelectric (TE) material for near room temperature applications and plays a significant role in TE technology.[1,2] Recently, much attention has been paid to the study of $Sb_2Te_3$ and related other layered stoichiometric compounds for the realization of a new class of quantum matter and topological insulator (TI).[3,4] A TI has a band gap in the bulk, however gapless Dirac like surface states are protected by time-reversal symmetry. These topologically protected surface states hold great promise for potential applications, including field effect transistors, infra-red (IR)-terahertz (THz) detectors, magnetic field sensors, quantum computations and others.[1,3,5] Apart from being TE material and its new role as TIs, $Sb_2Te_3$ is utilized in chalcogenide alloys as the phase changing materials for information storage.[6] Although $Sb_2Te_3$ and related other compounds like $Bi_2Se_3$, $Bi_2Te_3$ and others are being extensively studied, it is important to characterize them for their temperature dependent physical properties more accurately for device designing. It is also essential to understand the scattering mechanism, role of defects and study their effect on structural and transport properties in order to achieve the best device performance.

The efficiency of a TE material is evaluated by the term figure of merit (*ZT*), a dimensionless quantity, defined as $ZT = \frac{S^2}{\rho \kappa} T$, where *S*, $\rho$ and $\kappa$ are Seebeck coefficient, electrical resistivity, and thermal conductivity of the TE material, respectively.[7] An efficient TE material should possess high *ZT*. To improve the *ZT* value, high power-factor ($PF = \frac{S^2}{\rho}$) and low $\kappa$ must be obtained simultaneously.

Sb$_2$Te$_3$ is a layered compound which crystallizes in rhombohedral phase having symmetry space group $R\overline{3}m - D_{3d}^5$ with two fold and three fold axes of symmetry $C_2$ and $C_3$, respectively. Periodically ordered layers, in the plane perpendicular to the $C_3$ axis form the crystal lattice of Sb$_2$Te$_3$. Each layer consists of five atomic planes ordered in the following sequence: Te$^1$-Sb-Te$^2$-Sb-Te$^1$, where Te$^1$ and Te$^2$ denote the positions of Te atoms. Each of the five atomic planes is referred to as a quintuple (QL) layer and a weak Van-der-Waals interaction exists between the adjacent layers.[8,3] This weak Van-der-Waals interaction between the QLs leads to the interesting physical properties in Sb$_2$Te$_3$ and related materials. The weak interlayer interaction usually involves antisite (AS) defects, i.e., Sb atoms in Te position in Sb$_2$Te$_3$. Due to such native defects, Sb$_2$Te$_3$ always shows a *p*-type conductivity with hole concentration up to $10^{20}$ cm$^{-3}$.[8,9] Often it has been reported that the temperature dependent $\rho$ ($\rho(T)$) data depicts metallic behavior, i.e., the value of $\rho$ is increased with the increase of temperature in Sb$_2$Te$_3$ system.[3,9-11] Dutta *et al.*[3] pointed out the role of surface states for such metallic $\rho(T)$ data, observed in bulk Sb$_2$Te$_3$ and related TIs. However, it may be pointed out that Sb$_2$Te$_3$ is a narrow band gap (E$_g$) semiconductor. The E$_g$ in Sb$_2$Te$_3$, appearing because of the energy separation between the top of upper valence band (UVB) and bottom of lower conduction band (LCB), is reported to be around 200 meV.[10,12] Direct measurements by tunnelling experiments reveal that E$_g$ = 238 meV at room temperature and the value increases to around 260 meV at 4.2 K.[13]

For further understanding of the TE property of Sb$_2$Te$_3$ based alloy, we have synthesized both metallic and semiconducting Sb$_2$Te$_3$ samples by tuning the preparation condition. Here, we report the detailed structural analysis, temperature dependent transport properties, viz, $\rho(T)$ and temperature dependent S ($S(T)$), of both metallic and semiconducting Sb$_2$Te$_3$ samples. We have also specifically emphasised on the estimation of PF value for the synthesized samples and a comparison has been made for their TE properties. Further, an attempt has been made to elucidate the origin of metallic and semiconducting $Sb_2Te_3$ and find out its correlation with the estimated TE property.

## II. EXPERIMENTAL DETAILS

Polycrystalline $Sb_2Te_3$ samples were synthesized by solid state reaction method. Two types of $Sb_2Te_3$ samples, *viz.*, metallic (S1) and semiconducting (S2), were prepared by tuning the cooling rate of the melted mixture. Stoichiometric amount of Sb and Te (each of purity 99.999%; Alfa Aesar, UK) were sealed in quartz tube under pressure $10^{-3}$ Pa to avoid the oxidation. Vacuum sealed quartz ampoules were initially annealed at temperature 1123 K for 24 h. It is then sintered at 893 K for 96 h to homogenize the alloys, followed by liquid Nitrogen quenching. S1 was cooled from 1123 K to 893 K at 5 K/h, whereas S2 was cooled at 8 K/h in the same temperature range.

The structural characterization of the synthesized $Sb_2Te_3$ alloys were carried out using powder X-ray diffractometer (Model: X'Pert Powder, PANalytical) with Cu-$K_\alpha$ radiation of wavelength 0.15418 nm. All the X-ray diffraction (XRD) measurements were performed on powdered samples in the range of $10^0 \leq 2\theta \leq 120^0$ in θ-2θ geometry. Rietveld refinement technique, utilizing *Materials Analysis Using Diffraction* (MAUD) program, was used to perform in depth structural analysis of the samples.[14,15] Standard Si was used to determine the instrumental profile.[16] $\rho(T)$ was measured using conventional four probe method down to 10 K. Electrical contacts were made by silver paste, cured at room temperature. $S(T)$ of the synthesized samples, down to 20 K was measured using standard differential technique. A small temperature gradient ($\Delta T$) was created across the sample and the voltage ($\Delta E$) developed between the hot and cold end was measured. At a particular temperature, the thermopower was calculated from the slope of ΔE versus Δ$T$ plot to eliminate any contribution from spurious emf. The $\rho(T)$ and $S(T)$ experiments were carried out on parallelepiped shaped samples of dimensions around 6 x 4 x 0.8 mm$^3$. Room temperature Raman spectroscopic (inVia, Renishaw) studies were performed in the range of 50 to 400 cm$^{-1}$ using 514.5 nm of Ar$^+$ laser using small flakes of respective samples with sub-micron focusing diameter (objective of 50X magnification with numerical aperture 0.8). An 1800 gr/mm grating was used for monochromatization with thermoelectric cooled charged coupled device (CCD) as detector in the back scattering configuration.

## III. RESULTS AND DISCUSSION

XRD patterns recorded for both the synthesized samples, S1 and S2, are shown in Fig. 1. The XRD spectra reveal that the samples are single phase in nature without the presence of any impurity phase, at least within the detectable limit of XRD. All the diffraction peaks are indexed with rhombohedral phase. As compared to S1, full width at half maxima (FWHM) of the diffraction peaks of S2 is broader and the corresponding peak intensities are smaller (Fig. 1). The broadening of diffraction peaks and lower intensity implies the inferior crystalline quality of the semiconducting sample. Broadening of XRD peaks is also associated with the presence of defects. Increased number of defect density in S2, as also evidenced from Raman spectroscopy, resistivity and thermopower data (discussed later), might lead to the observed inferior crystalline quality of S2. In addition, Rietveld refinement using the program MAUD[15] was also employed to perform in-depth structural analysis of the samples and the XRD patterns after refinement is also included in Fig. 1. We have performed the refinement using the atomic positions and substitutions of the synthesized $Sb_2Te_3$ samples. Space group $R\bar{3}m$ and point group $D_{3d}$ were used for the refinement. The extracted unit cell parameters along with the corresponding goodness of fit (GoF) or $\chi^2$ values and other refinement parameters like site occupancy, atomic positions, reliability parameters ($R_w$, $R_b$, $R_{exp}$) along with other parameters are provided in Table I. It should be pointed out that the microstructural refinement according to the formalism of Popa,[17] is capable of modeling both isotropic and anisotropic size and strain broadening. This has been used in the present Rietveld analysis. In order to confirm the isotropic/anisotropic nature of the synthesized samples, we performed Williamson-Hall (W-H) analysis.[18,19] The non-linearity of the experimental data points in the W-H plot confirms that the microstructure of both the $Sb_2T_3$ samples are anisotropic i.e. direction dependent and hence refined according to the Popa model incorporated in Rietveld refinement code MAUD.[17] Here the volume-weighted crystallite sizes/particle sizes ($D_v$) along <006>, as obtained from Rietveld analysis with anisotropic line broadening scheme is considerably smaller (11 nm) with correspondingly higher values of strain in S2 in comparison with S1 (21 nm). On the other hand, it lies between 9 – 12 nm in other directions for S2 and 13-19 nm for S1 (Table I). The anisotropic size-strain analysis clearly reveals that the estimated grain size along individual planes is significantly smaller with correspondingly higher value of strain in S2. Further the Debye-Waller factors ($B_{iso}$) or temperature factors for both S1 and S2 are quite high (Table I). One of the inherent reasons for this is the high background scattering and fluctuations in it. However, for the $Sb_2Te_3$ system AS defect is quite common.[8,9] Sb atoms may occupy off-centered position because of difference in ionic sizes of Sb and Te. This kind of off-centering leads to the departure from rhombohedral structure locally, giving rise to high $B_{iso}$ values.[20]

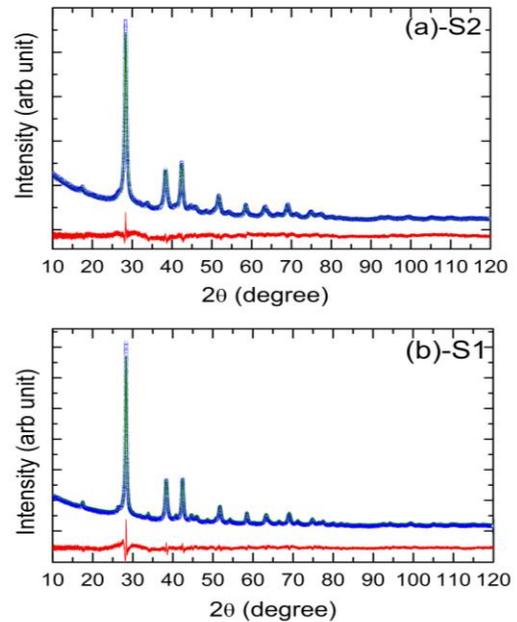

FIG. 1. X-ray diffraction patterns of Sb$_2$Te$_3$ alloy: (a) semiconducting (S2) and (b) metallic (S1) after Rietveld refinement.

Room temperature Raman spectrum (RS) for S1 and S2 are displayed in Fig. 2. It is noteworthy to mention that Raman spectroscopy is an excellent tool for providing useful structural information related to crystalline phase and stoichiometry of the samples.[1] As mentioned earlier, $Sb_2Te_3$ exhibits a layered rhombohedral crystal structure of space group $R\bar{3}m$. Each rhombohedral unit cell contains five atoms and consequently $Sb_2Te_3$ exhibits fifteen normal modes at the $\Gamma$ point of the Brillouin zone represented by the irreducible representation: $\Gamma=2(A_{1g}+E_g)+3(A_{2u}+E_u)$, where $g$ and $u$ represent the active Raman and IR modes, respectively.[21,22] Three of these modes are acoustic and twelve are optical phonons. Out of the twelve optical phonon modes, four modes e.g., $E_g^1$, $E_g^2$, $A_{1g}^1$, $A_{1g}^2$ are Raman active in the frequency range 30-200 $cm^{-1}$.[5] The $A_{1g}^1$ and $E_g^1$ modes occur at lower frequencies than the $A_{1g}^2$ and $E_g^2$ modes.[1] Sosso $et$ $al.$[21] reported the following active Raman modes for $Sb_2Te_3$ at room temperature and ambient pressure, $E_g$ modes at 46 and 113 $cm^{-1}$ and $A_{1g}$ modes at 62 and 166 $cm^{-1}$.

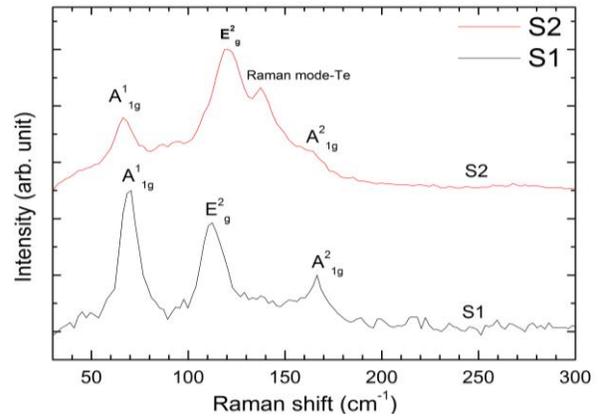

FIG. 2. Room temperature Raman spectra of S2 and S1 samples indicating the Raman active $A_{1g}^1$, $E_g^2$, $A_{1g}^2$ modes of $Sb_2Te_3$ recorded under excitation at λ = 514.5 nm. In addition, Raman spectrum of S2 shows presence of Raman active mode of Tellurium at 135.5 $cm^{-1}$.

The RS of sample S1 (Fig. 2) clearly indicates the presence of three peaks centred at about 69, 112 and 166.6 $cm^{-1}$. The position of the peaks are determined by fitting Lorentzian function and compared with reported Raman active modes.[21-23] The 112 cm$^{-1}$ peak is attributed to Raman active $E_g^2$ mode, while the peaks at 69 and 166.6 cm$^{-1}$ are attributed to Raman active $A_{1g}^1$ and $A_{1g}^2$ modes, respectively. The active Raman mode $E_g^1$ at 46 cm$^{-1}$ is out of the range measured in this work. On the other hand, RS of sample S2 (Fig. 2) exhibits three peaks centred at about 66.3, 120, and 135.5 cm$^{-1}$ along with a broad low intensity peak at around 165 cm$^{-1}$. The appearance of the new low intensity peak at 135.5 cm$^{-1}$ in S2 overlaps with 165 cm$^{-1}$ peak, making the later as a broad hump. Like RS of S1, all the reported Raman active modes of Sb$_2$Te$_3$ in the measured experimental range viz., $E_g^2$ (120 cm$^{-1}$), $A_{1g}^1$ and $A_{1g}^2$ (66.3 and 165 cm$^{-1}$, respectively) are also clearly identified for S2 (Fig. 2). However, the peak centred at 135.5 cm$^{-1}$ is not associated with Sb$_2$Te$_3$. Identification of this peak is performed by comparing the spectrum with the reported RS of Sb, Te, Sb$_2$O$_3$ and TeO$_2$.[24] The data reported by Torrie *et al.*[25] unambiguously indicates that the 135.5 cm$^{-1}$ peak can be attributed to Raman active mode of Te. Therefore RS reflects minute amount of Te segregation in semiconducting sample and thus S2 shows a microstructure formed of a matrix of Sb$_2$Te$_3$ and segregated Te particles. Depending on annealing condition, partial segregation of elemental Te during the synthesis of Sb$_2$Te$_3$ has also been reported by Abrikosov *et al.*[26] Since the diffraction peaks of crystalline Te, being present in small amounts, are not observed in the XRD pattern (Fig. 1). As shown in the XRD data, Fig. 2 also divulge that all the Raman active peaks of S2 are broadened as compared to S1. Broadening of Raman peaks is associated with the presence of defects, and amorphization.[27] Thus, in corroboration with XRD and Raman data it is quite justified to assume that the defect density in semiconducting Sb$_2$Te$_3$ is higher than that for the metallic one. The defects in semiconducting Sb$_2$Te$_3$ (S2) sample is mostly Te vacancy mediated point defects or its agglomeration. This is further supported from the density of the samples, estimated using site occupancy and unit cell volume obtained from Rietveld Refinement data.[18] Density of S2 sample (6.521 ± 9.45 x $10^{-3}$ g $cm^{-3}$) is slightly lower as compared to S1 sample (6.551 ± 0.18 x $10^{-3}$ g $cm^{-3}$), due to the presence of Te vacancy.[28] The site occupancy parameter, as obtained from Rietveld refinement data, depicts that $Te^2$ site in Te sublattice has vacancy of around 2.7% (Table I). Here it may be pointed out that the saturation vapor pressure of Sb and Te is 10 and 1000 Pa, respectively, at around 890 K (homogenization temperature).[29] Since S2 is cooled at a faster rate than that for S1, volatilization of Te atom may play a dominant role in introducing Te vacancies.[30] Moreover, Te vacancies ($V_{Te}$) are easier to be generated than Sb vacancies ($V_{Sb}$), because of higher saturation vapor pressure in the former. Thus, Te may have segregated in minute amount in the matrix of S2 sample.

The $\rho$–$T$ curve of the polycrystalline $Sb_2Te_3$ samples are shown in Fig. 3. Sample S1 depicts metallic behaviour, i.e., the value of $\rho$ increases with increasing temperature. Sample S2 shows semiconducting nature, where the value of $\rho$ decreases with increasing temperature. Thus, the $\rho(T)$ curves, measured in the temperature range of 300 to 10 K, clearly indicate metallic and semiconducting nature of the synthesized $Sb_2Te_3$ samples. Figure 3 also reveals that the value of $\rho$ for S2 is higher than that of S1. The observed increased value of $\rho$ in S2 is intimately related to the increased impurity scattering. It is quite plausible to assume that defects, viz., $V_{Te}$ as well as segregated elemental Te in the matrix of semiconducting $Sb_2Te_3$ (S2) act as scattering sites. Thus $\rho(T)$ also substantiate the XRD and Raman spectroscopic results, indicating that semiconducting $Sb_2Te_3$ sample (S2) have higher defect density. $Sb_2Te_3$ is a narrow band gap semiconductor. Near the band edge the energy spectrum of $Sb_2Te_3$ has two valence bands; a light hole band or UVB and a heavy hole band or lower valence band (LVB). There are also two conduction bands, an upper conduction band (UCB) and LCB.[8] This semiconducting nature of $Sb_2Te_3$ is correctly reflected in the $\rho(T)$ data of S2. In addition, $\rho(T)$ curve of S1 depicts good metallic behavior (Fig. 3). In accordance to some of the previous reports on $Sb_2Te_3$, the value of $\rho(T)$ for S1 is also comparable to that of $Bi_2Se_3$ and $Bi_2Te_3$ crystals with carrier density around $10^{18}$-$10^{19}$ cm$^{-3}$.[3,9,10,31-33] This is typical for heavily doped semiconductors and quite natural for $Sb_2Te_3$ system, for which high hole concentration (~$10^{20}$ cm$^{-3}$) is already reported.[3,9,10] Blank et al.[10] explained the metallic behavior by the increase of the intrinsic carrier concentration at high temperatures in narrow band semiconductors. Excessive carrier concentration in $Sb_2Te_3$ based TI is originated from inherent AS defects. The Fermi energy, $E_F$ goes deep inside the bulk valence band resulting in metallic nature of the sample.[34] However, in comparison to the above mentioned reports for similar systems, the residual resistivity ratio ($\rho_{300K}/\rho_{10K}$) of ~1.25 is little less than the earlier reports.[3,31-33] It should be mentioned that most of the reports are on single crystalline materials, whereas the present study is based on polycrystalline materials, where grain boundary plays a significant role. In the temperature region of 10-300 K, the $\rho(T)$ curve of S1 is fitted with the power-law expression $\rho=\rho_o+AT^n$. Boltzmann transport mechanism predicts, $\rho_{eph}=AT^n$, where $\rho_{eph}$ is the resistivity for electron-electron (e-e) or electron-phonon (e-ph) interaction. The best fit value of $n$ for S1 is 1.66. Similar value of $n$ for $Sb_2Te_3$ crystals was also reported by Dutta et al.[3]

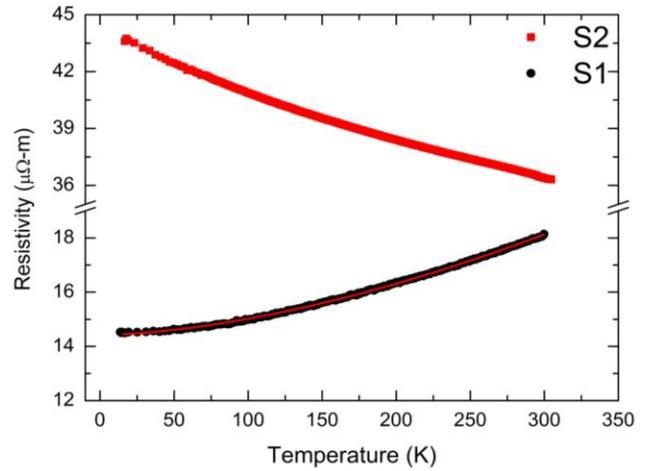

FIG. 3. Thermal variation of resistivity ($\rho$) of S2 and S1. Solid line indicates the power-law fit, $\rho = \rho_o + AT^n$ with n ~ 1.66 to the $\rho$(T) data for sample S1.

In general, at low temperature $\rho(T)$ follows $T^2$ dependence for pure metal, where the value of $\rho$ arises mostly because of the $e$-$e$ interaction.[35] In magnetic materials, $\rho$ shows $T^{3/2}$ dependence in the low temperature regime because of the magnetic contribution to scattering. On the other hand, we demonstrated the contribution of both $e$-$e$ ($T^2$) and $e$-$ph$ ($T^5$) scattering in resistivity for Bi-Sb alloy based TE materials.[36] Thus, the obtained best fit value of 1.66 for the exponent does not classify this compound to hither-to-known conventional metallic system. As mentioned above, $Sb_2Te_3$ is a well-known TI. For bulk samples of TIs, transport properties of the surface state are often mixed with the bulk state, which probably gives rise to the unusual value of the exponent in the metallic $Sb_2Te_3$ (S1) sample.[3] Temperature dependent thermopower (Seebeck Coefficient, $S$) of S1 and S2 are shown in Fig. 4. The $S$ of both the samples depicts positive value, which increases with increasing temperature. The positive value of $S$ indicates hole transport, i.e., both the samples are $p$-type in nature.[37] According to Abrikosov et al.[26], the $Sb_2Te_3$ crystals prepared from the melt of stoichiometric composition 2Sb/3Te exhibit always a surplus of Sb. The underlying reason is that Te partially segregates during the growth.[26] The over-stoichiometric Sb atoms occupy, prevailingly the Te sites in the Te sublattice giving rise to AS defects. The $p$-type conductivity observed in $Sb_2Te_3$ arises as a natural consequence of AS defects. The room temperature $S$ values of S1 and S2 are about 60 and 115 µV/K, respectively. For $Sb_2Te_3$, similar $S$ value is also reported by other research groups.[37-40] Further, the value of $S$ for the semiconducting sample (S2) is higher than that for S1. It is noteworthy to mention that the value of $\rho$ for S2 is also higher. It is realized that S2 hosts larger density of defect sites than that of S1. These defects act as scattering centre, enhancing the carrier scattering. This in a turn is reflected in the increased values of $\rho$ and $S$ for sample S2.

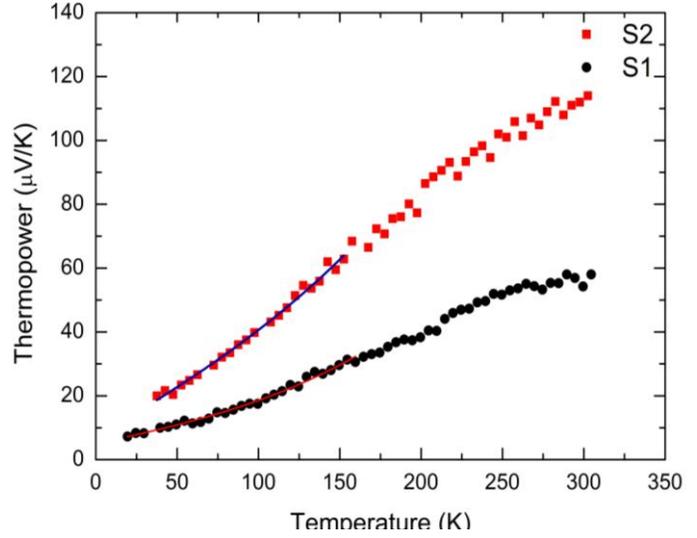

FIG. 4. Temperature dependence of thermopower (S) of semiconducting (S2) and metallic (S1) $Sb_2Te_3$. Solid lines indicate the best fit with equation $S = AT + BT^3$.

Thus, due to the presence of different scattering sites, e.g., AS defect in both S1 and S2, $V_{Te}$ type point defect or segregated Te predominantly in S2, different scattering mechanisms significantly influence the $S(T)$ data. In order to extract their contribution, the corresponding $S(T)$ data below Debye temperature ($\theta_D$) have been fitted with the equation,

$$S = AT + BT^3 \qquad (1)$$

where, $A = \dfrac{\pi^2 k_B^2}{3qE_F}$ and $B = \dfrac{4\pi^4 k_B^4}{5q\theta_D}$. Here $q$ is the charge of the carrier and $k_B$ is the Boltzmann constant. It should be mentioned that $AT$ is the contribution from diffusion thermopower, where carrier scattering plays a significant role. In the second term, B is the coefficient of carrier-phonon (c-ph) interaction which is dependent on $T^3$.[36,41,42] Barnard explicitly demonstrated that the carrier-carrier scattering and carrier-impurity atom scattering mostly contribute in diffusion thermopower.[42] The value of $\theta_D$ for binary alloys can be calculated from the Kopp-Neumann relation,

$$\theta_D^{-3} = (1-x)\theta_1^{-3} + x\theta_2^{-3}, \qquad (2)$$

where $\theta_1$ and $\theta_2$ are the Debye temperatures of the Sb and Te, respectively.[43] The estimated $\theta_D$ of $Sb_2Te_3$ is 154 K.[44]

The best fit curve of equation (1) with $S(T)$ data for both the samples is included in Fig. 4 and the corresponding best fit values of $A$ and $B$ are provided in Table II. It is evidenced that both the scattering coefficients, $A$ and $B$ are higher for semiconducting $Sb_2Te_3$ (S2) than those for metallic one (S1). The higher defect density of S2 acts as scattering center and affects the thermal transport of carriers in S2 considerably, thus giving rise to higher value of estimated scattering coefficients. Furthermore, it should be pointed out that the scattering coefficient $A$ $(=\frac{\pi^2 k_B^2}{3qE_F})$ is inversely proportional to $E_F$. The semiconducting $Sb_2Te_3$ (S2) sample is having inherent Te vacancy ($V_{Te}$) related point defects. The presence of vacancy related point defect leads to the depletion of carrier density in the system, which further lowers the $E_F$. The lowering of $E_F$, thus correctly leads to the increased value of observed scattering coefficient $A$ for S2.[45] Increase in the measured values of $S$ and/or $\rho$ with decreasing carrier density for $Sb_2T_3$ based system have already been reported by Draser et al.[39] and Lostak et al.[40] In addition, c-ph scattering coefficient ($B$) is also higher for S2 than that of S1. Phonon scattering plays a dominant role in an alloy because of the reason that the random point defects, created due to alloying, act as ideal scattering sites. It is worth mentioning that the contribution due to c-ph interaction dominates, as the temperature is increased.[36,41,46,47] It has been shown for doped metal di-silicides alloys and skutterudite compounds like In-filled $CoSb_3$, the lower lattice thermal conductivity or better thermoelectric performance is likely due to point defect scattering and c-ph scattering, which increases with increasing temperature.[45,47] Also recently, Varshney et al.[46] has demonstrated that contribution due to c-ph increases with temperature and improves the thermopower value of polycrystalline manganites at higher temperature. The evidence of enhanced phonon scattering was also experimentally verified on SiGe based thermoelectric materials.[48] Phonon scattering is an effective route leading to the reduction of lattice thermal conductivity and better performance in thermoelectric materials. However, it should be clearly mentioned that several efforts have been made to study the effect of scattering on thermal conductivity and $ZT$.[49,50] But not much attention has been given to reveal the effect of scattering on PF, which is also an important factor in controlling the efficiency of a TE material.

The calculated thermal variation of PF for S1 and S2 are displayed in Fig. 5. The PF is calculated from experimental values of $S$ and $\rho$. The PF values of both the samples increase monotonically with increasing temperature. Moreover, the value of PF obtained here for semiconducting sample (S2) is higher than that of its metallic counterpart (S1) in the whole temperature range investigated. It has already been established that the defect density is higher in S2. These defect states, predominantly $V_{Te}$ as well as segregated Te increase disorder in the crystal lattice and scatter the charge carriers along with phonons so that both the values of $\rho$ and $S$ increase for S2. The enhancement in the thermopower, $S$ outweighs the increase in $\rho$, which causes the PF of S2 to increase (Fig. 5). The higher value of PF for S2 than that of S1 suggests that the $p$-type semiconducting $Sb_2Te_3$ will possess better TE performance. However, conclusive evidence should be drawn only after measuring thermal conductivity and estimating $ZT$ of the present $Sb_2Te_3$ samples. Similar defect state induced enhancement of TE performance has recently demonstrated by others also.[45,51,52] However, while explaining the thermopower data it has been realised that the S2 sample either shows depletion of carrier (hole) density or it has reduced amount of acceptor concentration. Blank *et al.*[10] also demonstrated that the performance of TE nano-composites can be significantly enhanced by the reduction of acceptor concentration.

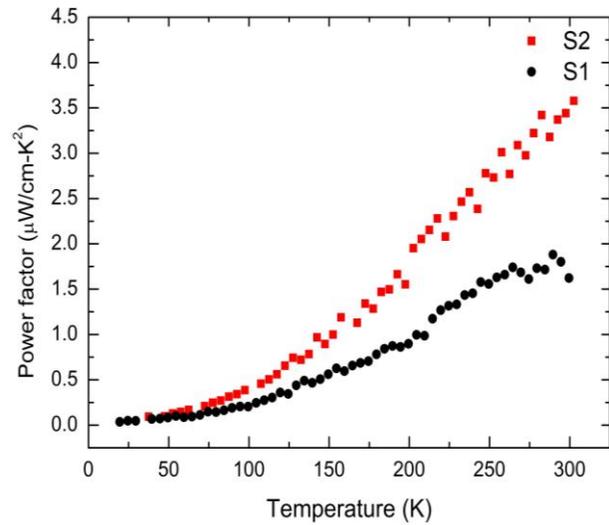

FIG. 5. Temperature dependent power factor of semiconducting (S2) and metallic (S1) $Sb_2Te_3$.

## IV. CONCLUSION

Polycrystalline metallic and semiconducting $Sb_2Te_3$ samples are synthesized by tuning the annealing condition. A higher value of strain in semiconducting $Sb_2Te_3$ in comparison with its metallic counterpart is analyzed from the XRD studies in details using Popa model incorporated in Rietveld refinement code MAUD. Raman spectroscopic studies also supported the presence of defect in the semiconducting phase from the observations of Raman active mode of Te and broadening of the symmetry allowed vibrational modes. Te vacancy ($V_{Te}$) is formed due to higher saturation vapor pressure of Te and it leads to the segregation in minute amount in the matrix of semiconducting $Sb_2Te_3$ sample. Temperature dependent resistivity, $\rho(T)$ plots also confirm the presence of metallic and semiconducting phases in these samples. The $\rho(T)$ curve of metallic $Sb_2Te_3$ sample is fitted with the power-law expression $\rho = \rho_o + AT^n$, with an unusual value of the exponent, $n=1.66$ indicating presences of an unconventional metallic phase in the sample. It has been realized that the surface states are often mixed with the bulk state, giving rise to the observed metallicity in $Sb_2Te_3$. The temperature dependence of thermopower, $S(T)$ data for both the samples increases monotonically with increasing temperature and the positive value of it indicates *p*-type nature of the synthesized samples. The $\rho(T)$, $S(T)$ as well as temperature dependence of power factor (PF) indicate higher value of $\rho$, $S$ and PF in semiconducting sample than those for the metallic sample. Semiconducting $Sb_2Te_3$ hosts larger numbers of defects, predominantly $V_{Te}$ type point defect and segregated Te in $Sb_2Te_3$ matrix, which act as scattering center giving rise to increased value of $\rho$, $S$ and PF. The role of different scattering mechanisms on the transport as well as the thermoelectric properties of $Sb_2Te_3$ has been discussed and it is concluded that the sample with higher carrier and phonon scattering strength possess higher PF. We have clearly demonstrated that the incorporation of defects in $Sb_2Te_3$ system represents a good opportunity for improving its thermoelectric performance.

**Acknowledgment:**

This work is supported by Department of Science and Technology (DST), Govt. of India and UGC-DAE CSR, Kalpakkam node in the form of sanctioning research project, reference no. SR/FTP/PS-25/2009 and CSR-KN/CRS-65/2014-15/505. Authors KM and DD are thankful to UGC, Govt. of India and UGC-DAE CSR, Govt. of India, respectively for providing research fellowships.

**TABLE I.** Rietveld refinement parameters, viz., lattice parameters, unit cell volume, grain size and strain, positional coordinates, site occupancy, Debye-Waller factor ($B_{iso}$), sample height offset parameter ($Z_{disp}$), reliability parameters [$R_w$(%), $R_b$(%), $R_{exp}$ (%)] and goodness of fit (Gof or $\chi^2$) value, as obtained using MAUD software for semiconducting and metallic $Sb_2Te_3$ samples. The corresponding values of the estimated errors are also given.

| Phase | $Sb_2Te_3$ [$R\bar{3}m$] (Semiconducting) | $Sb_2Te_3$ [$R\bar{3}m$] (Metallic) |
|---|---|---|
| Cell (Å) | $a$: 4.2568 $\pm$ 5.7 x 10$^{-5}$<br>$c$: 30.3193 $\pm$ 1.0 x 10$^{-3}$ | $a$: 4.2558 $\pm$ 4.3 x 10$^{-5}$<br>$c$: 30.3629 $\pm$ 7.4 x 10$^{-4}$ |
| Cell Vol. (Å$^3$) | 475.79 | 476.25 |
| Size in nm (hkl) | 11(006),12(015),9(1010),12(110),10(205), 12(0210), 10(1115),12(125), 10(2110) | 21(006),19(015),14(1010),16(110),13(205), 19(0210), 16(1115),17(125), 14(2110) |
| Strain x 10$^{-3}$ (hkl) | 6.4 (006), 1.2x10$^{-2}$ (015), 2.7(1010),2.7(110), 3.2(205), 0.31(0210), 2.4(1115), 2.4(125), 2.9(2110) | 5.2 (006), 6.0 x10$^{-4}$ (015), 2.8 (1010),1.5(110), 1.5(205), 8.1 x10$^{-2}$ (0210), 2.4(1115),1.5(125), 1.3(2110) |
| $Sb_x$/$Sb_y$/$Sb_z$ | 0.0 / 0.0/ 0.3992 $\pm$ 4.7x10$^{-5}$ | 0.0 / 0.0/ 0.3994 $\pm$ 4.1x10$^{-5}$ |
| $Te1_x$/$Te1_y$/$Te1_z$ | 0.0 / 0.0/ 0.7874 $\pm$ 2.6x10$^{-5}$ | 0.0 / 0.0/ 0.7874 $\pm$ 2.4x10$^{-5}$ |
| $Te2_x$/$Te2_y$/$Te2_z$ | 0.0/ 0.0/ 0.0 | 0.0/ 0.0/ 0.0 |
| $B_{iso\ Sb/Te1/Te2}$ | Sb: 2.686 $\pm$ 0.029<br>Te1: 1.452 $\pm$ 0.035<br>Te2: 0.130 $\pm$ 0.030 | Sb: 2.619 $\pm$ 0.026<br>Te1: 1.522 $\pm$ 0.032<br>Te2: 0.147 $\pm$ 0.029 |
| Site occupancy Sb / Te1/ Te2 | 1.0 / 1.0 / 0.973 $\pm$ 4.3x10$^{-3}$ | 1.0 / 1.0 / 1.0 |
| $Z_{disp.}$ | 0.2612 $\pm$ 0.0021 | 0.2544 $\pm$ 0.0013 |
| $R_w$ (%) | 3.690 | 3.941 |
| $R_b$(%) | 2.872 | 3.091 |
| $R_{exp}$(%) | 2.083 | 2.104 |
| Gof | 1.771 | 1.873 |

**TABLE II**. The best fit value of the parameters A and B; as obtained from fitting thermopower data with equation: $S = AT + BT^3$.

| Sample | A ($\mu VK^{-2}$) | B ($\mu VK^{-4}$) |
| --- | --- | --- |
| S1 (Metallic) | 1.129 x $10^{-7}$ | 2.2998 x $10^{-12}$ |
| S2 (Semiconducting) | 3.0968 x $10^{-7}$ | 2.7703 x $10^{-12}$ |